\documentclass[journal]{IEEEtran}
\usepackage{caption}
\usepackage{graphicx}
\usepackage{comment}
\usepackage{tikz}

\usepackage{pgfplots,pgfplotstable}
\usetikzlibrary{pgfplots.groupplots}
\pgfplotsset{compat=1.5}

\usepackage{subcaption} 

\usepackage{multirow}
\usepackage{booktabs}
\usepackage{makecell}

\usepackage{amsmath}
\usepackage{amsthm}

\usepackage{textcomp}
\usetikzlibrary{shapes,arrows}

\usepackage{empheq} 

\usepackage{stfloats} 

\usepackage[hyphens]{url}
\usepackage[hidelinks]{hyperref}
\hypersetup{colorlinks=true,linkcolor=black,citecolor=black,filecolor=black,urlcolor=black,breaklinks=true}
\usepackage{cite} 

\usepackage{caption} 
\usepackage{color}

\usepackage{cancel} 

\usepackage{array} 

\usepackage[export]{adjustbox} 

\usepackage[normalem]{ulem} 

\usepackage{tabularx}
\usepackage{ragged2e}
\newcolumntype{Y}{>{\RaggedRight\arraybackslash}X}

\usepackage{amsfonts} 

\usepackage{cite}
\usepackage[hyphens]{url}
\usepackage[hidelinks]{hyperref}
\hypersetup{colorlinks=true,linkcolor=black,citecolor=black,filecolor=black,urlcolor=black,breaklinks=true}
\usepackage{float}
\usepackage{color}
\usepackage{amsmath,amssymb,amsfonts}
\usepackage{algorithmic}
\usepackage{graphicx}
\usepackage{svg}
\usepackage{amssymb} 
\usepackage{bbding} 
\usepackage{balance}
\usepackage{textcomp}
\usepackage{xcolor}
\usepackage{multirow}
\usepackage{multicol}
\def\BibTeX{{\rm B\kern-.05em{\sc i\kern-.025em b}\kern-.08em
    T\kern-.1667em\lower.7ex\hbox{E}\kern-.125emX}}

\usepackage{booktabs}
\usepackage{tabularx}
\usepackage{array}
\usepackage{ragged2e}
\newcolumntype{L}[1]{>{\RaggedRight\arraybackslash}p{#1}}
\newcolumntype{C}[1]{>{\centering\arraybackslash}p{#1}}

\makeatletter
\def\bstctlcite{\@ifnextchar[{\@bstctlcite}{\@bstctlcite[@auxout]}}
\def\@bstctlcite[#1]#2{\@bsphack
 \@for\@citeb:=#2\do{%
   \edef\@citeb{\expandafter\@firstofone\@citeb}%
   \if@filesw\immediate\write\csname #1\endcsname{\string\citation{\@citeb}}\fi}%
 \@esphack}
\makeatother

\begin{document}

\title{Voltage Sensitivity-Based Fair Allocation of Operating Envelopes in Distribution Networks}

\author{Alireza~Zabihi,~\IEEEmembership{Student Member,~IEEE,}
        Maurizio~Vassallo,~\IEEEmembership{Student Member,~IEEE,}
        Damien~Ernst,
        Luis~Badesa,
        and~Araceli~Hernández,~\IEEEmembership{Senior Member,~IEEE} 
}

\maketitle

\begin{abstract}
Operating envelopes (OEs) define the admissible active and reactive power injections of distributed energy resources while respecting distribution network limits. In unbalanced low-voltage feeders, OEs based only on voltage magnitude and thermal limits may overestimate available flexibility because they neglect power quality requirements. This paper develops a three-phase unbalanced AC optimal power flow framework to compute coupled active and reactive power OEs while enforcing limits on the voltage unbalance factor (VUF). It also proposes a fairness mechanism based on voltage sensitivities to allocate available flexibility among multiple photovoltaic (PV) units. The proposed mechanism is compared with lexicographic max--min fairness as a benchmark for equitable resource allocation. Case studies on unbalanced test feeders evaluate how VUF constraints change the feasible operating region of PV units. The results show how VUF constraints affect feasible flexibility and demonstrate that the proposed voltage sensitivity-based mechanism is nearly six times faster than lexicographic max--min fairness.
\end{abstract}


\IEEEpeerreviewmaketitle

\section*{Nomenclature}
\addcontentsline{toc}{section}{Nomenclature}

\subsection*{Indices and Sets}
\begin{IEEEdescription}[\IEEEusemathlabelsep\IEEEsetlabelwidth{$\mathcal{F}^{(s)}_\phi,\,\mathcal{B}^{(s)}_\phi$}]
\item[$i,\,k,\,\mathcal{N}$]
   Indices of network nodes; set of all network nodes, $|\mathcal{N}|=N$.
\item[$s$]
   Stage index in the lexicographic max-min procedure.
\item[$u,\,v$]
   Indices of DER units.
\item[$\phi,\,\mu,\,\Phi$]
   Phase indices; set of phases $\{a,b,c\}$.
\item[$\mathcal{L}$]
   Set of network branches.
\item[$\mathcal{U}_\phi$]
   Set of nodes hosting a DER on phase $\phi$.
\item[$\mathcal{U}$]
   Set of all DER units, $\mathcal{U}=\bigcup\limits_{\phi\in\Phi}\mathcal{U}_\phi$.
\item[$\mathcal{F}$]
   AC feasible set defined by network and operational constraints.
\item[$\mathcal{F}^{(s)}_\phi,\,\mathcal{B}^{(s)}_\phi$]
   Free and bound unit sets at stage $s$ of the lexicographic procedure on phase $\phi$.
\item[$\mathcal{E}_u$]
   Operating envelope polygon of unit $u$ in the P--Q plane.
\end{IEEEdescription}

\subsection*{Network Parameters}
\begin{IEEEdescription}[\IEEEusemathlabelsep\IEEEsetlabelwidth{$G_{\phi\phi,ik}$}]
\item[$\mathbf{Y}$]
    Three-phase nodal admittance matrix,
    $\mathbf{Y}\in\mathbb{C}^{3N\times 3N}$.
\item[$\mathbf{Z}$]
    Nodal impedance matrix, $\mathbf{Z}=\mathbf{Y}^{-1}$.
\item[$\mathbf{z}^{(u)}$]
    Column of $\mathbf{Z}$ corresponding to unit $u$, $\mathbf{z}^{(u)}\in\mathbb{C}^{3N}$.
\item[$Z_{ik}$]
    Branch impedance of line $(i,k)\in\mathcal{L}$.
\item[$Y^\mathrm{sh}_i$]
    Shunt admittance at node $i$.
\item[$G_{\phi\mu,ik}$]
    Real part of the element corresponding to phase $\phi$ at node $i$ and phase $\mu$ at node $k$ in the network admittance matrix $\mathbf{Y}$.
\item[$B_{\phi\mu,ik}$]
    Imaginary part of the element corresponding to phase $\phi$ at node $i$ and phase $\mu$ at node $k$ in the network admittance matrix $\mathbf{Y}$.
\item[$\ell_u$]
    Global column index of unit $u$ in $\mathbf{Z}$,
    $\ell_u = 3(i_u - 1) + \phi_u$.
\end{IEEEdescription}

\subsection*{Variables}
\begin{IEEEdescription}[\IEEEusemathlabelsep\IEEEsetlabelwidth{$P^\mathrm{DG}_{\phi,i},\,Q^\mathrm{DG}_{\phi,i}$}]
\item[$V^r_{\phi,i},\,V^i_{\phi,i}$]
    Real and imaginary parts of the phase-$\phi$ voltage
    phasor at node $i$.
\item[$V^r, V^i$] 
   Vectors of real and imaginary nodal voltage coordinates, $\in \mathbb{R}^{3N}$.
\item[$\mathbf{v}_{\phi,i}$]
    Complex voltage phasor,
    $\mathbf{v}_{\phi,i}=V^r_{\phi,i}+jV^i_{\phi,i}$.
\item[$V^{(0)}_u$]
    Baseline complex voltage at the bus of unit $u$.
\item[$V^+_i,\,V^-_i$]
    Positive- and negative-sequence voltage phasors
    at node $i$.
\item[$P^G_{\phi,i},\,Q^G_{\phi,i}$]
    Real and reactive power output of the conventional
    generator at node $i$, phase $\phi$.
\item[$P^G, Q^G$] 
   Vectors of real and reactive power outputs of conventional generators, $\in \mathbb{R}^{3N}$.
\item[$P^\mathrm{DG}_{\phi,i},\,Q^\mathrm{DG}_{\phi,i}$]
    Real and reactive power injection of the DER
    at node $i$, phase $\phi$.
\item[$P^\mathrm{DG}, Q^\mathrm{DG}$]
   Vectors of real and reactive power injections of DER units, $\in \mathbb{R}^{3N}$.
\item[$\mathbf{x}$] 
   Decision vector of the AC-OPF problem, comprising the collection of all voltage and power injection vectors ($\mathbf{x} = \{ {V}^r, {V}^i, {P}^G, {Q}^G, {P}^\mathrm{DG}, {Q}^\mathrm{DG} \}$).
\item[$\alpha$]
    Shared headroom scalar optimized at each angle
    and stage of fairness methods.
\item[$\alpha^{*(s)}_m$]
    Optimal value of $\alpha$ at stage $s$ and
    angle $\theta_m$.
\item[$\alpha_{u,m}$]
    Final allocated headroom scalar of unit $u$
    at angle $\theta_m$.
\end{IEEEdescription}

\subsection*{Constants and Parameters}
\begin{IEEEdescription}[\IEEEusemathlabelsep\IEEEsetlabelwidth{$P^D_{\phi,i},\,Q^D_{\phi,i}$}]
\item[$c_i$]
    Linear generation cost coefficient at node $i$.
\item[$P^D_{\phi,i},\,Q^D_{\phi,i}$]
    Active and reactive load demand at node $i$,
    phase $\phi$.
\item[$\hat{P}_u,\,\hat{Q}_u$]
    Baseline real and reactive injection of unit $u$
    from the baseline OPF solution.
\item[$\underline{P}^G_i,\,\overline{P}^G_i$]
    Lower and upper real power limits of the
    conventional generator at node $i$.
\item[$\underline{Q}^G_i,\,\overline{Q}^G_i$]
    Lower and upper reactive power limits of the
    conventional generator at node $i$.
\item[$\overline{S}^\mathrm{DG}_{\phi,i}$]
    Apparent power capacity of the DER at node $i$,
    phase $\phi$.
\item[$\underline{V}_i,\,\overline{V}_i$]
    Lower and upper voltage magnitude limits at
    node $i$.
\item[$\varepsilon$]
    Admissible VUF limit (e.g.\ $\varepsilon=2\%$
    per IEC~61000-2-2).
\item[$\theta_m$]
    $m$-th angular sample for OE boundary tracing,
    $\theta_m = 2\pi m/M$.
\item[$M$]
    Number of angular samples per OE boundary trace.
\item[$\omega$]
    Symmetrical components rotation operator,
    $\omega = e^{j2\pi/3}$.
\end{IEEEdescription}

\subsection*{Sensitivity and Fairness Quantities}
\begin{IEEEdescription}[\IEEEusemathlabelsep\IEEEsetlabelwidth{$P^\mathrm{OE}_{u,m},\,Q^\mathrm{OE}_{u,m}$}]

\item[$\sigma_u$]
    Aggregate voltage impact of unit $u$,
    $\sigma_u = \|\partial|\mathbf{V}|/\partial P_u\|^2
               +\|\partial|\mathbf{V}|/\partial Q_u\|^2$.
\item[$\tilde{w}_u$]
    Raw inverse-sensitivity weight of unit $u$,
    $\tilde{w}_u = 1/(\sqrt{\sigma_u}+\varepsilon_0)$.
\item[$w_u$]
    Phase-normalised sensitivity weight of unit $u$,
    $w_u\in(0,1]$.
\item[$\delta_u$]
    Step magnitude of unit $u$ from its baseline,
    $\delta_u=\|(\Delta P_u, \Delta Q_u)\|$.
\item[$P^\mathrm{OE}_{u,m},\,Q^\mathrm{OE}_{u,m}$]
    Real and reactive components of the $m$-th
    boundary point of the OE
    of unit $u$.
\item[$U^{*}_{\phi,m}$]
    Maximum aggregate step of the units on phase $\phi$
    at angle $\theta_m$ (utilitarian benchmark).
\item[$A^{F}_{u}$]
    Area of the OE polygon of unit $u$ under method
    $F\in\{\mathrm{M1},\mathrm{M2}\}$, where U
    denotes the utilitarian benchmark.
\item[$\mathrm{PoF}^{F}$]
    Price of fairness of method $F$.
\end{IEEEdescription}

\section{Introduction}
High penetrations of distributed energy resources (DERs), particularly inverter-interfaced photovoltaic (PV) generation, are reshaping the operational paradigm of medium- and low-voltage distribution networks. The resulting bidirectional and highly time-varying power flows can trigger voltage magnitude excursions and thermal congestion, thereby motivating network-aware mechanisms that communicate permissible DER operating limits without requiring the distribution system operator (DSO) to directly dispatch behind-the-meter devices. In this context, operating envelopes (OEs) have emerged as a practical interface between DSOs and DER owners/aggregators by specifying feasible import/export ranges that preserve network integrity \cite{liuOchoaOE,wickramasingheDOEReview,alamDOEAllocation}. Compared with static connection limits designed for worst-case conditions, the dynamic assignment of OEs can unlock latent hosting capacity by adapting limits to prevailing operating states, forecasts, and controllable network assets \cite{liuOchoaOE,arenaDOEReport,aemoEDGE, vassallo2025seculex}.

A large body of OE research adopts optimization-based computation, most commonly via optimal power flow (OPF), to embed voltage and thermal constraints into the envelope definition \cite{liuOchoaOE,alamDOEAllocation,wickramasingheDOEReview}. Because distribution feeders are inherently unbalanced due to single- and two-phase laterals, untransposed lines, and unevenly distributed single-phase loads and DER connections, three-phase unbalanced modeling is often essential for envelope accuracy and for preventing phase-specific violations \cite{liuOchoaOE,giraldoLinearACOPF,Chen2020Aggregate}. Recent advances have further expanded envelope concepts beyond scalar import/export limits to multi-dimensional feasible regions in the active--reactive (P--Q) plane, enabling explicit utilization of inverter reactive power capability for voltage management and service provision \cite{gaoEquitablePQEnvelope,Bruno2021Mapping}. Nevertheless, computation and allocation of P--Q envelopes remain challenging because they must balance network security, inverter capability constraints, and the equitable sharing of scarce network capacity among multiple DERs \cite{wickramasingheDOEReview,gaoEquitablePQEnvelope,Churkin2026Exposing}.

In parallel with voltage magnitude and thermal constraints, power quality constraints—most notably voltage unbalance (VU)—are receiving renewed attention. Increased single-phase DER uptake and uneven phase loading can exacerbate VU, which may elevate losses and adversely impact three-phase equipment \cite{girigoudarVUImpact,ZABIHI-J1}. Multiple standards define and quantify VU differently (e.g., IEC, IEEE, and NEMA), and the associated metrics are not fully consistent, which complicates the use of VU in optimization-based operation \cite{pillayDefinitionsVU,ZABIHI-C2}. In practice, the IEC definition based on symmetrical components is widely used through the voltage unbalance factor (VUF), i.e., the ratio of negative- to positive-sequence voltage magnitudes \cite{iec61000_3_13, zabihi_VUACOPF}. Importantly, OPF-based coordination that neglects VUF constraints can inadvertently aggravate unbalance even when voltage magnitude limits are satisfied, motivating the explicit inclusion of VUF constraints in distribution optimization \cite{fanKockarVUF,girigoudarVUImpact}.

Another critical aspect of OE deployment is fairness. Envelopes effectively allocate a limited network resource, namely hosting capacity, among many DERs. Purely efficiency-driven allocations can therefore produce systematically unequal curtailment or flexibility access, undermining social acceptance and regulatory feasibility \cite{wickramasingheDOEReview,aemoEDGE}. Existing OE allocation approaches have therefore begun to incorporate equitable perspectives and to analyze fairness properties, including proportional-fair allocations under uncertainty and explicit equitable allocation strategies \cite{liuBraslavskyRDOE,alamDOEAllocation}. 

The joint consideration of P--Q flexibility, VUF constraints, and fairness in OE allocation remains underexplored \cite{russell-Thesis}. This gap is important because VUF constraints can change the size and shape of the feasible P--Q region, while the choice of fairness mechanism determines how the remaining flexibility is distributed among DERs. To address this gap, this paper develops a unified framework for computing VUF-aware P--Q operating envelopes for PV inverters in unbalanced three-phase distribution networks. The framework uses a three-phase unbalanced AC-OPF formulation with nodal VUF constraints and inverter capability limits. A directional OPF sweep traces the boundary of the feasible P--Q region.

Existing studies have investigated proportional fairness and utility-based allocation mechanisms to balance efficiency and equity in flexibility allocation \cite{binObaidFairnessChapter,liuBraslavskyRDOE,Petrou2020,Moret2019}. Lexicographic max--min fairness offers an alternative that prioritizes the least-favored participant through a hierarchical allocation procedure \cite{ogryczakLexiMaxMin,binObaidFairnessChapter,Bertsekas1987,Huo2022}. Building on these approaches, this paper develops a network-aware proportional fairness mechanism based on voltage sensitivities with respect to power injections and compares it with lexicographic max--min fairness as the main reference of fairness mechanisms.

The main contributions of this paper are as follows:
\begin{itemize}
    \item A framework for computing coupled P--Q operating envelopes in unbalanced three-phase distribution networks using a directional OPF sweep, with explicit enforcement of nodal VUF constraints and PV inverter capability limits.

    \item A network-aware proportional fairness mechanism that allocates available flexibility among PV units based on voltage sensitivities to power injections, balancing allocation efficiency and equity.

    \item A comparative assessment of proportional fairness and lexicographic max--min fairness on two unbalanced test feeders. The assessment quantifies the effects of VUF constraints on OE size and shape, evaluates the price of fairness of each mechanism relative to a utilitarian benchmark, and examines the operational outcomes and flexibility allocation resulting from the two fairness mechanisms.
\end{itemize}

Table~\ref{tab:soa_clean} positions the proposed framework relative to representative studies on operating envelopes and voltage unbalance.

\begin{table*}[t]
\captionsetup{justification=centering, textfont={sc,footnotesize}, labelfont=footnotesize, labelsep=newline}
\caption{Positioning of this paper relative to representative OE and VU literature}
\label{tab:soa_clean}
\centering
\footnotesize
\setlength{\tabcolsep}{5pt}
\renewcommand{\arraystretch}{1.12}
\begin{tabularx}{\textwidth}{L{3.6cm} C{2.9cm} C{1.8cm} C{1.6cm} Y}
\toprule
\textbf{Reference} &
\textbf{Network model} &
\textbf{P--Q} &
\textbf{VUF} &
\textbf{Fairness / allocation} \\
\midrule
Liu \emph{et al.} \cite{liuOchoaOE} &
3$\phi$, linear {OPF} &
P-only &
No &
Not emphasized \\
\addlinespace[1pt]

Alam \emph{et al.} \cite{alamDOEAllocation} &
3$\phi$, linear {OPF} &
P-only &
No &
Technical vs.\ (soft-)equitable \\
\addlinespace[1pt]

Liu \& Braslavsky \cite{liuBraslavskyRDOE} &
3$\phi$, linear robust {OPF} &
P-only &
No &
Proportional fairness property (implicit) \\
\addlinespace[1pt]

Gao \emph{et al.} \cite{gaoEquitablePQEnvelope} &
3$\phi$, linear {OPF} &
\textbf{Yes} &
No &
Equitable allocation \\
\addlinespace[1pt]

Fan \& Kockar \cite{fanKockarVUF} &
\textbf{3$\phi$, {AC} {OPF}} &
-- &
\textbf{Yes} &
Not emphasized \\
\addlinespace[1pt]

Russell \emph{et al.} \cite{russellISGT2023} &
3$\phi$, linear robust {OPF} &
P-only &
\textbf{Yes} &
Not emphasized \\

Chen \emph{et al.} \cite{Chen2020Aggregate} &
\textbf{3$\phi$, {AC} {OPF}} &
\textbf{Yes} &
No &
Not emphasized \\
\addlinespace[1pt]

Churkin \emph{et al.} \cite{Churkin2026Exposing} &
\textbf{3$\phi$, {AC} {OPF}} &
\textbf{Yes} &
\textbf{Yes} &
Not emphasized \\
\addlinespace[1pt]

\midrule
\textbf{This paper} &
\textbf{3$\phi$, {AC} {OPF}} &
\textbf{Yes} &
\textbf{Yes} &
\textbf{Proposed voltage sensitivity-based proportional allocation; lexicographic max–min benchmark} \\
\bottomrule
\end{tabularx}
\end{table*}

The remainder of the paper is structured as follows: Section~\ref{sec:oe_full} introduces the three-phase VUF-constrained AC-OPF model, explains the envelope tracing procedure, elaborates on the fairness formulations, and defines the price-of-fairness metric. Section~\ref{sec:case_studies} presents the case studies along with comparative results. Section~\ref{sec:discussion} offers a discussion and comparative analysis of the findings; and finally, Section~\ref{sec:conclusion} summarizes and concludes the paper.

\section{Fair Operating Envelope Computation} 
\label{sec:oe_full}
This section formulates a three-phase unbalanced AC-OPF framework for computing coupled P--Q operating envelopes under explicit VUF constraints, and it embeds two alternative fairness mechanisms for allocating feasible flexibility across multiple DERs. 

\subsection{Network Model and Problem Setting}
\label{sec:notation}

Consider a three-phase unbalanced radial distribution network with
node set $\mathcal{N} = \{1,\dots,N\}$ and branch set $\mathcal{L}$.
Node~1 is the slack bus.
Voltages are expressed in rectangular coordinates
$\mathbf{v}_{\phi,i} = V^r_{\phi,i} + jV^i_{\phi,i} \in \mathbb{C}$
for phase $\phi \in \Phi = \{a,b,c\}$ and node
$i \in \mathcal{N}$.
The three-phase nodal admittance matrix
$\mathbf{Y} \in \mathbb{C}^{3N \times 3N}$ is assembled from
branch impedances $Z_{ij}$ and shunt admittances $Y^\mathrm{sh}_i$.

Let $\mathcal{U}_\phi \subseteq \mathcal{N}$ denote the set of nodes
hosting a DER on phase $\phi$ and
$\mathcal{U}=\bigcup_{\phi\in\Phi}\mathcal{U}_\phi$.
For each unit $u \in \mathcal{U}_\phi$, let $i_u$ be its bus index
and $(\hat{P}_u, \hat{Q}_u)$ its baseline real and reactive injection.
The $\mathcal{E}_u$ of unit $u$ is a polygon
in the P--Q plane whose boundary is traced at $M$ angular samples
$\theta_m = 2\pi m/M$, $m = \{0,\dots,M-1$\} representing its OE.

\subsection{Baseline Three-Phase AC Optimal Power Flow}
\label{sec:baseline_opf}

The baseline dispatch determines the optimal decision vector 
$\mathbf{x} = \{ {V}^r, {V}^i, {P}^G, {Q}^G, {P}^\mathrm{DG}, {Q}^\mathrm{DG} \}$, 
comprising nodal voltage coordinates in rectangular form and active/reactive power injections for both conventional generation and DERs. 
The objective is to minimize the total cost of conventional generation (e.g., feeder bus imports/exports):
\begin{equation}
  \min_{\mathbf{x}} \quad
  \sum_{i \in \mathcal{N}} \sum_{\phi \in \Phi} c_i P^G_{\phi,i}
  \label{eq:obj_baseline}
\end{equation}
where $c_i$ represents the cost coefficient assigned to conventional sources. The optimization is subject to the following constraints for all $i \in \mathcal{N}$ and $\phi \in \Phi$:

\paragraph{Power balance}
The nodal injections must satisfy the three-phase AC power flow equations for each phase:
\begin{align}
  & P^G_{\phi,i} + P^\mathrm{DG}_{\phi,i} - P^D_{\phi,i} = \sum_{k \in \mathcal{N}} \sum_{\mu \in \Phi} \Bigl[ \nonumber \\
  &\quad V^r_{\phi,i} \bigl( G_{\phi\mu,ik} V^r_{\mu,k} - B_{\phi\mu,ik} V^i_{\mu,k} \bigr) \nonumber \\
  &\quad + V^i_{\phi,i} \bigl( G_{\phi\mu,ik} V^i_{\mu,k} + B_{\phi\mu,ik} V^r_{\mu,k} \bigr) \Bigr], \label{eq:pbal} \\
  & Q^G_{\phi,i} + Q^\mathrm{DG}_{\phi,i} - Q^D_{\phi,i} = \sum_{k \in \mathcal{N}} \sum_{\mu \in \Phi} \Bigl[ \nonumber \\
  &\quad V^i_{\phi,i} \bigl( G_{\phi\mu,ik} V^r_{\mu,k} - B_{\phi\mu,ik} V^i_{\mu,k} \bigr) \nonumber \\
  &\quad - V^r_{\phi,i} \bigl( G_{\phi\mu,ik} V^i_{\mu,k} + B_{\phi\mu,ik} V^r_{\mu,k} \bigr) \Bigr]\,. \label{eq:qbal}
\end{align}

\paragraph{Voltage magnitude bounds}
Nodal voltages must remain within regulatory limits to ensure network stability:
\begin{equation}
  \underline{V}_i^2 \;\leq\;
  \bigl(V^r_{\phi,i}\bigr)^2 + \bigl(V^i_{\phi,i}\bigr)^2
  \;\leq\; \overline{V}_i^2\,.
  \label{eq:vmag}
\end{equation}

\paragraph{Generation limits}
\begin{equation}
  \underline{P}^G_i \leq P^G_{\phi,i} \leq \overline{P}^G_i\,,
  \qquad
  \underline{Q}^G_i \leq Q^G_{\phi,i} \leq \overline{Q}^G_i\,.
  \label{eq:gen_bounds}
\end{equation}

\paragraph{DER apparent power capacity}
\begin{equation}
  \bigl(P^\mathrm{DG}_{\phi,i}\bigr)^2 +
  \bigl(Q^\mathrm{DG}_{\phi,i}\bigr)^2
  \leq
  \bigl(\overline{S}^\mathrm{DG}_{\phi,i}\bigr)^2\,.
  \label{eq:solar_cap}
\end{equation}

\paragraph{Voltage unbalance factor constraint}
VU is quantified using the symmetrical component definition, i.e., the ratio of negative- to positive-sequence voltage magnitudes \cite{iec61000_3_13}. Using $\omega=e^{j2\pi/3}$, the positive- and negative-sequence phasors at bus $i$ are:
\begin{align}
  V^-_i &= \tfrac{1}{3}\!\left(\mathbf{v}_{a,i} + \omega^2\mathbf{v}_{b,i} + \omega\mathbf{v}_{c,i}\right)\,, \label{eq:vneg}\\
  V^+_i &= \tfrac{1}{3}\!\left(\mathbf{v}_{a,i} + \omega\mathbf{v}_{b,i} + \omega^2\mathbf{v}_{c,i}\right)\,. \label{eq:vpos}
\end{align}
The nodal VUF limit is enforced via:
\begin{equation}
  |V^-_i|^2 \le \varepsilon^2 |V^+_i|^2\,, \qquad \forall i\in\mathcal{N}\,,
  \label{eq:vuf}
\end{equation}
where $\varepsilon$ denotes the admissible VUF threshold (e.g., $\varepsilon=0.02$ corresponds to 2\% VUF).
Let $\mathcal{F}$ denote the feasible set defined by \eqref{eq:pbal}--\eqref{eq:vuf}.
Note that \eqref{eq:vuf} is quadratic in the voltage variables and, when coupled with the nonlinear three-phase AC power-flow constraints, yields a nonconvex nonlinear program for OE tracing.

\subsection{Voltage Sensitivity Weights}
\label{sec:sensitivity}
In this section, the foundations of weight calculations for proportional fairness are presented.
Let $\mathbf{Z} = \mathbf{Y}^{-1} \in \mathbb{C}^{3N \times 3N}$ denote the system impedance matrix. For each unit $u \in \mathcal{U}_\phi$, we define its global column index $\ell_u$ as:
\begin{equation}
    \ell_u = 3(i_u - 1) + \phi_u
\end{equation}
where $i_u$ is the node index and $\phi_u \in \{1, 2, 3\}$ represents the integer 
phase index of unit $u$, so that $\mathbf{z}^{(u)} \triangleq \mathbf{Z}_{:,\,\ell_u}$ 
denotes the $\ell_u$-th column of $\mathbf{Z}$.

Linearizing the bus voltage equation $\mathbf{V} = \mathbf{Z}\mathbf{I}$
around the baseline and converting via
$\Delta I_u \approx \Delta S_u^*/\overline{V}^{(0)}_u$
gives the Z-bus voltage sensitivity~\cite{Tinney,Talkington}:
\begin{align}
  \frac{\partial|\mathbf{V}|}{\partial P_u}
    &\approx \operatorname{Re}\!\left(
       \frac{\mathbf{z}^{(u)}}{\overline{V}^{(0)}_u}
     \right)\,, \label{eq:dVdP}\\
  \frac{\partial|\mathbf{V}|}{\partial Q_u}
    &\approx \operatorname{Im}\!\left(
       \frac{\mathbf{z}^{(u)}}{\overline{V}^{(0)}_u}
     \right)\,. \label{eq:dVdQ}
\end{align}
The sensitivities~\eqref{eq:dVdP}--\eqref{eq:dVdQ} follow
from~\cite{Tinney,Talkington}. Based on these, we define an
aggregate voltage-impact score and a normalized weighting
scheme as follows.
The aggregate network-wide voltage impact of unit $u$ is:
\begin{equation}
  \sigma_u =
  \left\|\frac{\partial|\mathbf{V}|}{\partial P_u}\right\|^2 +
  \left\|\frac{\partial|\mathbf{V}|}{\partial Q_u}\right\|^2\,.
  \label{eq:sigma}
\end{equation}
The sensitivity weight is normalized within each phase group
independently:
\begin{equation}
  \tilde{w}_u = \frac{1}{\sqrt{\sigma_u}+\varepsilon_0}\,,
  \qquad
  w_u = \frac{\tilde{w}_u}
             {\displaystyle\max_{v \in \mathcal{U}_\phi}\tilde{w}_v}\,,
  \quad \forall\,u \in \mathcal{U}_\phi\,,
  \label{eq:weights}
\end{equation}
where $\varepsilon_0 > 0$ is a regularization constant.
Phase-local normalization ensures $w_u \in (0,1]$ with $w_u = 1$
assigned to the least voltage-sensitive unit on each phase,
so that allocation ratios reflect intra-phase network
heterogeneity rather than inter-phase impedance differences.

\subsection{Fair OE Computation Methods}
\label{sec:methods}
Having established the network model and the voltage sensitivity 
weights, two methods of computing fair allocated {OEs} that embed 
these ingredients are presented. Method~I pursues a closed-form 
proportional allocation, offering computational simplicity and 
transparency. Method~II formulates a lexicographic max--min 
optimization that prioritizes equity across units within each phase.

\subsubsection{Network-Weighted Proportional Fairness}
\label{sec:prop}
Method~I extends classical proportional fairness by weighting 
allocations with $w_u$, so that units with lower grid impact 
receive proportionally larger {OEs}. The resulting allocation 
is closed-form and computationally lightweight.

\paragraph{Formulation}
Proportional fairness assigns headroom according to each unit's 
ability to withstand loading variations in bus voltage, without 
breaching network constraints~\cite{Petrou2020,Moret2019}.
Here, the capacity is quantified by the sensitivity weight
$w_u \in (0,1]$.
For each phase $\phi$ and angle $\theta_m$, a single shared
scalar $\alpha \geq 0$ is maximized:
\begin{equation}
  \alpha^*_m = \max_{\alpha \geq 0,\;\mathbf{x} \in \mathcal{F}}
  \;\alpha
  \label{eq:prop_obj}
\end{equation}
subject to:
\begin{align}
  P^\mathrm{DG}_{\phi,i_u}
    &= \hat{P}_u + w_u\,\alpha\cos\theta_m\,,
  \quad \forall\,u \in \mathcal{U}_\phi\,,
  \label{eq:prop_p}\\
  Q^\mathrm{DG}_{\phi,i_u}
    &= \hat{Q}_u + w_u\,\alpha\sin\theta_m\,,
  \quad \forall\,u \in \mathcal{U}_\phi\,,
  \label{eq:prop_q}
\end{align}
and the full network feasibility constraints 
\eqref{eq:pbal}--\eqref{eq:vuf}.

This constitutes a single-level nonconvex nonlinear program in 
which \eqref{eq:prop_p}--\eqref{eq:prop_q} augment $\mathcal{F}$ 
with a shared proportionality structure across all units on phase $\phi$.
The boundary point of unit $u$ at angle $\theta_m$ is:
\begin{equation}
  \bigl(P^\mathrm{OE}_{u,m},\;Q^\mathrm{OE}_{u,m}\bigr)
  = \bigl(
      \hat{P}_u + w_u\alpha^*_m\cos\theta_m,\;
      \hat{Q}_u + w_u\alpha^*_m\sin\theta_m
    \bigr)\,.
  \label{eq:prop_boundary}
\end{equation}

\paragraph{Fairness Properties}
The allocation ratio between any two units on the same phase
is determined solely by their weights:
\begin{equation}
  \frac{\|\Delta\mathbf{s}_u\|}{\|\Delta\mathbf{s}_v\|}
  = \frac{w_u}{w_v}
  = \frac{\sqrt{\sigma_v}}{\sqrt{\sigma_u}} \,,
  \quad \forall\,u,v \in \mathcal{U}_\phi \,,
  \label{eq:prop_ratio}
\end{equation}
where $\Delta\mathbf{s}_u = (w_u\alpha^*_m\cos\theta_m,\,
w_u\alpha^*_m\sin\theta_m)$ denotes the shift of unit \(u\) relative to the
baseline.
This ratio is angle-independent and fully determined by the
linearized sensitivity~\eqref{eq:sigma} evaluated at the
baseline operating point.
The allocation is therefore Pareto-efficient and satisfies the
proportional fairness axioms of~\cite{Petrou2020}: no reallocation
can increase one unit's headroom by more than a proportional
factor without reducing another unit's by at least the same factor.

A structural limitation follows directly from~\eqref{eq:prop_ratio}:
the allocation ratios are fixed before the {OE} optimization is
solved and do not adapt to the actual feasible headroom at each
angle $\theta_m$.
Consequently, if the network constraint that limits $\alpha^*_m$
originates from a unit with large weight (electrically remote,
large $w$), a more sensitive unit with small $w$ may receive
less headroom than the feasible set $\mathcal{F}$ could support
for it, a conservatism that the method does not adapt within 
the current formulation.

\subsubsection{Lexicographic Max-Min Fairness}
\label{sec:lexmaxmin}
Lexicographic max-min fairness is a classical
fairness criterion in resource allocation and network
optimization~\cite{Bertsekas1987}.
It has recently been applied to fair allocation problems
in power and energy systems, including distribution-level
resource management and {EV} charging coordination~\cite{Sortomme2011,Huo2022}.

\paragraph{Fairness Properties}
A feasible allocation $\boldsymbol{\delta} = (\delta_1, \dots, \delta_n)$ is
lexicographically max-min optimal if and only if its sorted version
$\delta_{(1)} \leq \delta_{(2)} \leq \cdots \leq \delta_{(i)} \leq \cdots \leq \delta_{(n)}$
is lexicographically maximum over all feasible allocations: 
no reallocation can increase $\delta_{(i)}$ without
decreasing some $\delta_{(k)}$ with $k < i$.
This criterion implies both Pareto efficiency and proportional
fairness: any lexicographic max--min optimal allocation is also
Pareto efficient and proportionally fair, but the opposite does
not hold. The distinction is that the worst-off unit is
protected unconditionally, irrespective of gains provided to others.

\paragraph{Formulation}
For a fixed phase $\phi$ and angle $\theta_m$, the algorithm
proceeds through at most $|\mathcal{U}_\phi|$ sequential stages.
Let $\mathcal{F}^{(s)}_\phi$ and $\mathcal{B}^{(s)}_\phi$
denote the free and bound unit sets at stage $s$, initialised
as $\mathcal{F}^{(1)}_\phi = \mathcal{U}_\phi$ and
$\mathcal{B}^{(1)}_\phi = \emptyset$.

At stage $s$, the following optimization is solved:
\begin{equation}
  \alpha^{*(s)}_m =
  \max_{\alpha \geq 0,\;\mathbf{x} \in \mathcal{F}} \;\alpha
  \label{eq:lex_obj}
\end{equation}
subject to, the network feasibility constraints 
\eqref{eq:pbal}--\eqref{eq:vuf}, and for all free units $u \in \mathcal{F}^{(s)}_\phi$:
\begin{align}
  P^\mathrm{DG}_{\phi,i_u}
    &= \hat{P}_u + \alpha\cos\theta_m\,,
  \label{eq:lex_free_p}\\
  Q^\mathrm{DG}_{\phi,i_u}
    &= \hat{Q}_u + \alpha\sin\theta_m\,,
  \label{eq:lex_free_q}
\end{align}
and for all bound units $u \in \mathcal{B}^{(s)}_\phi$:
\begin{align}
  P^\mathrm{DG}_{\phi,i_u}
    &= \hat{P}_u + \alpha^{*(s_u)}_m\cos\theta_m\,,
  \label{eq:lex_fix_p}\\
  Q^\mathrm{DG}_{\phi,i_u}
    &= \hat{Q}_u + \alpha^{*(s_u)}_m\sin\theta_m\,,
  \label{eq:lex_fix_q}
\end{align}
where $s_u < s$ is the stage at which unit $u$ was fixed.
Every free unit moves by the same scalar $\alpha$ with no
external weight factor; the allocation ratios are not
predetermined but emerge from the feasibility of
$\mathcal{F}$ at each stage.

After solving, the binding unit set is identified as those
units whose {DER} apparent-power constraint~\eqref{eq:solar_cap}
is active at $\alpha^{*(s)}_m$:
\begin{equation}
\begin{aligned}
  \mathcal{B}^\mathrm{new}
  = \Bigl\{
    u \in \mathcal{F}^{(s)}_\phi \;&\Big|\;
    \overline{S}^2_{\phi,i_u}
    - \bigl(
      \hat{P}_u
      + \alpha^{*(s)}_m \cos\theta_m
    \bigr)^2 \\
    &-
    \bigl(
      \hat{Q}_u
      + \alpha^{*(s)}_m \sin\theta_m
    \bigr)^2
    \leq \tau
  \Bigr\}\,,
\end{aligned}
\label{eq:lex_bind}
\end{equation}
where $\tau > 0$ is a numerical tolerance.
These units are moved to $\mathcal{B}^{(s+1)}_\phi$ with
$\alpha_{u,m} = \alpha^{*(s)}_m$, and the procedure repeats
until $\mathcal{F}^{(s)}_\phi = \emptyset$.

The boundary point of each unit is:
\begin{equation}
  \bigl(P^\mathrm{OE}_{u,m},\;Q^\mathrm{OE}_{u,m}\bigr)
  = \bigl(
      \hat{P}_u + \alpha_{u,m}\cos\theta_m,\;
      \hat{Q}_u + \alpha_{u,m}\sin\theta_m
    \bigr)\,.
  \label{eq:lex_boundary}
\end{equation}

\subsection{Comparative Analysis}
\label{sec:comparison}

\subsubsection{Allocation structure}

The two methods share the same feasible set $\mathcal{F}$,
the same baseline $(\hat{P}_u, \hat{Q}_u)$, and the same
angular tracing framework.
They differ in one fundamental respect: in Method~I the
allocation ratios are fixed
by the linearized sensitivity~\eqref{eq:prop_ratio} and
do not depend on the angle $\theta_m$ or on which constraint
in $\mathcal{F}$ is active.
In Method~II the ratios vary with
$\theta_m$ and are determined by the nonlinear feasibility of
$\mathcal{F}$ at each stage.

This difference has a direct geometric interpretation.
In Method~I the OE polygon $\mathcal{E}_u$ is a scaled copy
of a common reference polygon, with the scaling factor $w_u$
fixed for all $u$.
In Method~II the shape of $\mathcal{E}_u$ varies across units
in a way that reflects the actual headroom available at each
location and direction: units that are electrically constrained
in one direction but not another will have asymmetric envelopes
that Method~I cannot reproduce.

\subsubsection{Sensitivity to linearisation accuracy}

Method~I depends critically on the accuracy of the Z-bus
sensitivity~\eqref{eq:dVdP}--\eqref{eq:dVdQ}, which is a
first-order linearisation around the baseline operating point.
If the true operating point deviates significantly from the
baseline, as occurs under high DER penetration or heavy loading, 
the weights $w_u$ become inaccurate and the allocation
ratios~\eqref{eq:prop_ratio} are systematically biased.
Method~II imposes no such linearisation: the sequential stage
solves are full AC-OPF problems over $\mathcal{F}$, so the
binding constraints are evaluated at the true nonlinear
operating point.
Method~II is therefore more robust to baseline deviation at
the cost of increased computational complexity.

\subsubsection{Fairness criteria satisfied}

Both methods satisfy Pareto-efficiency: no unit's envelope
can be enlarged without shrinking another's, because the
objective in both~\eqref{eq:prop_obj} and~\eqref{eq:lex_obj}
is the maximum feasible $\alpha$ over $\mathcal{F}$.
Method~I additionally satisfies the proportional fairness
axiom~\cite{Petrou2020}.
Method~II satisfies lex. max-min optimality, which is
logically independent of proportional fairness.

The summary of these conceptual comparisons is included in Table~\ref{tab:comparison}.

\begin{table*}[t]
\captionsetup{justification=centering, textfont={sc,footnotesize}, labelfont=footnotesize, labelsep=newline}
\caption{Comparison of fairness mechanisms used for P--Q operating-envelope allocation}
\label{tab:comparison}
\centering
\footnotesize
\setlength{\tabcolsep}{5pt}
\renewcommand{\arraystretch}{1.15}
\begin{tabularx}{\textwidth}{L{3.2cm} Y Y}
\toprule
\textbf{Aspect} &
\textbf{Method~I: Network-weighted proportional fairness} &
\textbf{Method~II: Lexicographic max--min fairness} \\
\midrule
Fairness criterion &
Proportional fairness via utility-based rationale \cite{Petrou2020,Moret2019} &
Lexicographic max--min optimality (worst-off protection) \cite{Huo2022,ogryczakLexiMaxMin} \\

Allocation structure &
Predefined ratios through network weights $w_u$; angle-independent scaling within a sweep direction &
Ratios emerge from feasibility; may vary with $\theta_m$ due to binding constraints \\

Required modeling ingredients &
Z-bus sensitivity for $w_u$ (linearization) + AC feasibility for OE tracing &
AC feasibility throughout; no sensitivity linearization required \\

OPF solves per angle and phase &
Single solve per $\theta_m$ and phase group &
Multiple staged solves; worst-case $\le |\mathcal{U}_\phi|$ per $\theta_m$ and phase group \\

Efficiency--equity trade-off &
May undermine equity and utilization&
Strongest guarantee for equity and utilization\\

Sensitivity to baseline point &
Dependent on linearization accuracy of $w_u$ around baseline &
Less dependent on baseline; constraints evaluated at nonlinear operating point \\
\bottomrule
\end{tabularx}
\end{table*}

\subsection{Price of Fairness}
\label{sec:pof}

To quantify the flexibility given up by each fairness
mechanism, a utilitarian benchmark is defined on the same
feasible set $\mathcal{F}$, the same baseline
$(\hat{P}_u,\hat{Q}_u)$, and the same angular samples
$\theta_m$. Each unit is assigned an individual step
$\delta_u$ along the common direction, and the aggregate
step is maximized:
\begin{equation}
  U^{*}_{\phi,m} =
  \max_{\delta_u \ge 0,\;\mathbf{x}\in\mathcal{F}} \;
  \sum_{u\in\mathcal{U}_\phi} \delta_u
  \label{eq:util}
\end{equation}
subject to \eqref{eq:pbal}--\eqref{eq:vuf} and, for all
$u \in \mathcal{U}_\phi$,
\begin{equation}
  P^\mathrm{DG}_{\phi,i_u} = \hat{P}_u + \delta_u \cos\theta_m\,,
  \qquad
  Q^\mathrm{DG}_{\phi,i_u} = \hat{Q}_u + \delta_u \sin\theta_m\,.
  \label{eq:util_dir}
\end{equation}
The allocations of Method~I ($\delta_u = w_u\alpha^{*}_m$) and
Method~II ($\delta_u = \alpha_{u,m}$) are both feasible points
of \eqref{eq:util}--\eqref{eq:util_dir}, so the benchmark
imposes no allocation rule beyond network feasibility.

The price of fairness of method
$F\in\{\mathrm{M1},\mathrm{M2}\}$ is defined as the relative
loss in total envelope area with respect to the benchmark:
\begin{equation}
  \mathrm{PoF}^{F} = 1 -
  \frac{\sum_{u\in\mathcal{U}} A^{F}_{u}}
       {\sum_{u\in\mathcal{U}} A^{\mathrm{U}}_{u}}\,,
  \label{eq:pof}
\end{equation}
where $A^{F}_u$ is the shoelace area of the polygon formed by
the boundary points
$\bigl(P^\mathrm{OE}_{u,m},Q^\mathrm{OE}_{u,m}\bigr)$ of unit
$u$ under method $F$. Since each boundary point lies on the ray
from the baseline in direction $\theta_m$, every polygon is
star-shaped about the baseline and its area is computed in
angular trace order. Envelope area is used because it is the
quantity that characterizes the OEs throughout this paper.
Since the area couples the steps of adjacent angles, maximizing
it directly would require solving all angles jointly; the
per-angle problem~\eqref{eq:util} is therefore used as a
tractable surrogate, and $\mathrm{PoF}^{F}\ge 0$ is verified
empirically rather than guaranteed by construction.

The simulation results in Section~\ref{sec:case_studies} quantify
these differences for two different networks under varying
DER penetration and loading levels.
The metrics reported are the OE polygon area per unit, 
the price of fairness~\eqref{eq:pof}, 
and the VUF constraint activity across the feeder.

\section{Test Cases and Simulation Setup}
\label{sec:case_studies}

This section introduces the two test systems, describes the
simulation setup, and presents the main numerical outcomes used in
the subsequent discussion. The first test system is a small
unbalanced feeder intended to illustrate the impact of embedding
VUF constraints in OE computation and to reveal the resulting
changes in the feasible P--Q regions at individual nodes. The
second test system is based on a simplified electrically
equivalent representation of the IEEE European LV feeder
\cite{Ahmad} and is used to evaluate the proposed framework under
a larger and more realistic setting, with particular attention to
the comparison between the two fairness inclusion methods.

\begin{figure}[!t]
    \centering
        \includegraphics[width=0.90\linewidth]{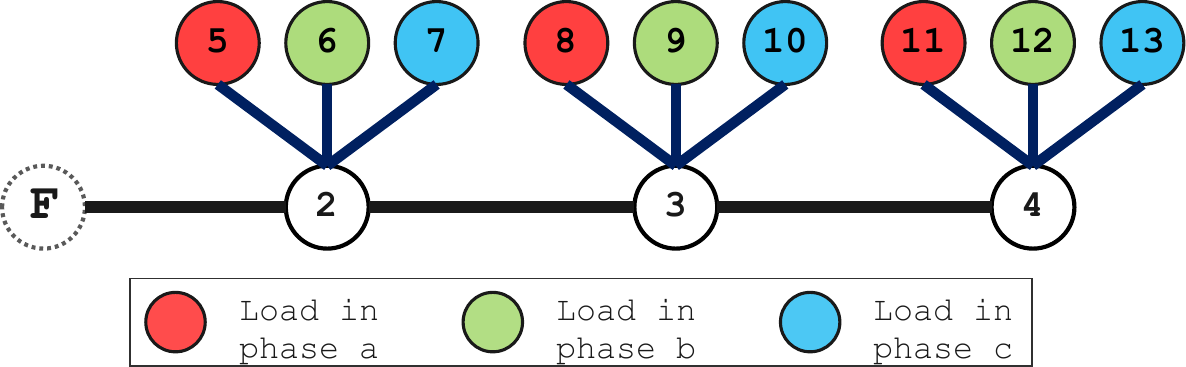}
    \caption{Network diagram for Test Case 1.}
    \label{fig:NW1}
\end{figure}

The first test case considers the 13-node network shown in
Fig.~\ref{fig:NW1}. The feeder includes nine nodes with
single-phase loads and rooftop PV units rated at 2.5~kVA. The total network demand is
18~kWh and the nominal PV capacity is 22.5~kVA. 

\begin{figure}[!t]
    \centering
        \includegraphics[width=1.0\linewidth]{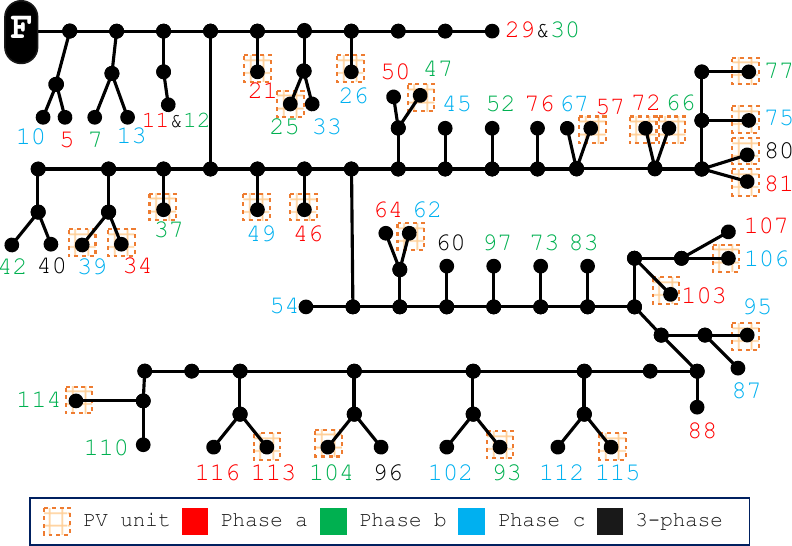}
    \caption{Network diagram for Test Case 2.}
    \label{fig:NW2}
\end{figure}

The second test case is based on the feeder described in \cite{Ahmad}, but modified to incorporate eight rooftop PV panel units per phase, each rated at 7.5~kVA, together with a three-phase solar generator rated at 22.5~kVA, and four three-phase loads representing a larger motor-type demand, resulting in a total feeder demand of 165~kWh. Fig.~\ref{fig:NW2} shows the network and modification applied on it.

In both test
systems, the phase loading follows the order $a>b>c$, so that the
network operates under intentionally asymmetric conditions. These
two cases provide complementary evidence: the first one highlights the
local and phase-specific effect of VUF-aware OE construction,
whereas the second one is better suited to assessing the scalability
and consistency of the two fairness formulations.

All simulations were implemented in \texttt{Julia} using
\texttt{JuMP} and solved with \texttt{Ipopt}. The simulations were
performed on a laptop with an Intel Core i7-1255U CPU at
1.70~GHz and 16~GB of RAM. The implementation is publicly
available online.\footnote{GitHub Repository:
\url{https://github.com/alireza33zz/OE_VU-F2-v1.0.git}} The
OEs were obtained through the directional OPF
sweep described in the Subsection~\ref{sec:methods}, first without and then
with explicit VUF constraints. Two fairness methods were
evaluated in each case: Method~I based on network-aware proportional fairness mechanism based on voltage sensitivities, and Method~II based on lexicographic max--min
fairness. 

From a computational perspective, Method~II is more intensive because it requires a sequence of solves for each angular direction, whereas Method~I needs only a single solve per direction. The runtimes in Table~\ref{tab:runtime} corroborate this behavior for both networks. For the smaller feeder NW1 (13 nodes, 9 DERs), Method~II is roughly 5$\times$ slower than Method~I, with runtimes of 56.14~s and 11.66~s, respectively. This disparity becomes even larger in the bigger feeder NW2 (116 nodes, 27 DERs), where Method~II takes 1274~s compared to 218~s for Method~I, nearly a 6$\times$ difference. This indicates that the scalability gap between the two methods grows as the network size and the number of DER units increase.
VUF enforcement introduces a steady but secondary computational overhead for both methods and both networks, leading to runtime increases between $\approx$5\% and $\approx$36\%. The only exception is Method~II in NW1, where a slight decrease in runtime is observed, which is probably attributed to a solver artifact.
\begin{table}[!t]
\captionsetup{justification=centering, textfont={sc,footnotesize}, labelfont=footnotesize, labelsep=newline}
    \centering
    \caption{Computational Time of Proposed Methods over Two Networks with $5^{\circ}$ (72 steps) Angular Resolution.}
    \label{tab:runtime}
    \renewcommand{\arraystretch}{1.35}
    \resizebox{\columnwidth}{!}{%
    \begin{tabular}{lccccc}
        \hline
        \makecell{\textbf{Network}\\\textbf{name}} & \makecell{\textbf{Number}\\\textbf{of nodes}} & \makecell{\textbf{Number}\\\textbf{of DERs}} & \textbf{Method} & \makecell{\textbf{without}\\\textbf{VUF (sec)}} & \makecell{\textbf{With}\\\textbf{VUF (sec)}} \\
        \hline
        \multirow{2}{*}{NW1} & \multirow{2}{*}{13} & \multirow{2}{*}{9} & M-I  & 11.7 & 12.3 \\
                             &                     &                     & M-II & 56.1 & 45.5 \\
        \hline
        \multirow{2}{*}{NW2} & \multirow{2}{*}{116} & \multirow{2}{*}{27} & M-I  & 217.5 & 296.1 \\
                             &                     &                     & M-II & 1273.5 & 1579.9 \\
        \hline
    \end{tabular}}
\end{table}

Table~\ref{tab:pof} reports the price of fairness~\eqref{eq:pof}
of both methods relative to the utilitarian
benchmark~\eqref{eq:util}.
Without VUF constraints, Method~II retains considerably more of
the benchmark envelope area than Method~I, with a PoF of 6.82\%
against 16.20\% in NW1 and 35.56\% against 43.61\% in NW2.
Enforcing VUF constraints raises the PoF of both methods, to
27.11\% (M1) and 24.89\% (M2) in NW1 and to 51.52\% (M1) and
54.21\% (M2) in NW2. At the same time, the difference between
the two methods narrows from 9.38 to 2.22 percentage points in
NW1 and from 8.05 to 2.69 percentage points in NW2, where
Method~I attains a slightly lower PoF than Method~II once VUF
constraints are active.

\begin{table}[!t]
\captionsetup{justification=centering, textfont={sc,footnotesize}, labelfont=footnotesize, labelsep=newline}
    \centering
    \caption{Price of Fairness of the Proposed Methods Relative to the Utilitarian Benchmark with $5^{\circ}$ (72 steps) Angular Resolution.}
    \label{tab:pof}
    \renewcommand{\arraystretch}{1.35}
    \begin{tabular}{lcccc}
        \hline
        \multirow{2}{*}{\makecell{\textbf{Network}\\\textbf{name}}} & \multicolumn{2}{c}{\textbf{Without VUF}} & \multicolumn{2}{c}{\textbf{With VUF}} \\
         & \textbf{M-I} & \textbf{M-II} & \textbf{M-I} & \textbf{M-II} \\
        \hline
        NW1 & 16.20\% & 6.82\%  & 27.11\% & 24.89\% \\
        NW2 & 43.61\% & 35.56\% & 51.52\% & 54.21\% \\
        \hline
    \end{tabular}
\end{table}

\begin{figure}[!t]
    \centering
        \includegraphics[width=1.0\linewidth]{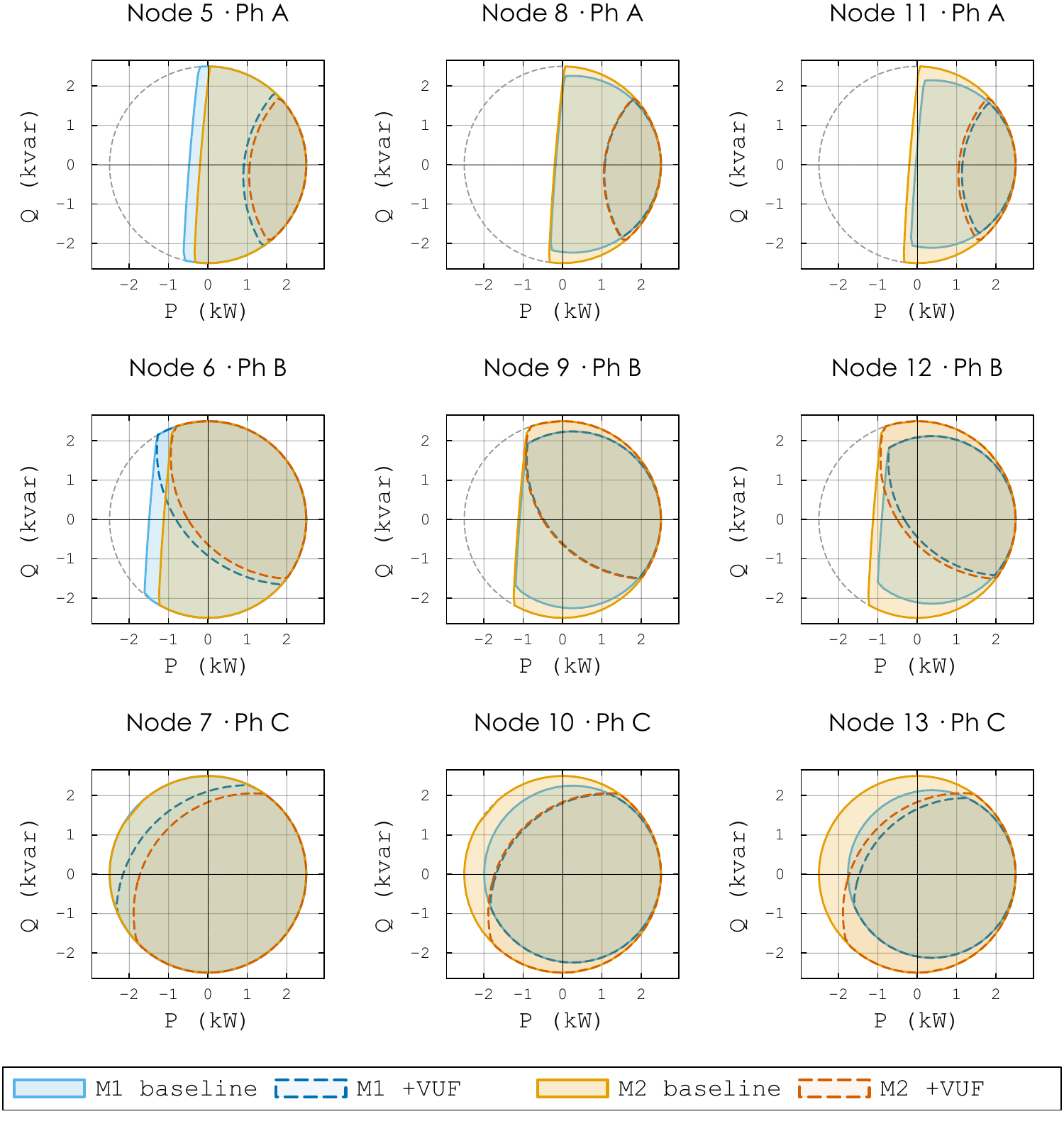}
    \caption{OE of Test Case 1 (Guide: M1:= Method~I, M2:= Method~II).}
    \label{fig:OE1}
\end{figure}

\begin{figure*}[!t]
    \centering
        \includegraphics[width=1.0\linewidth]{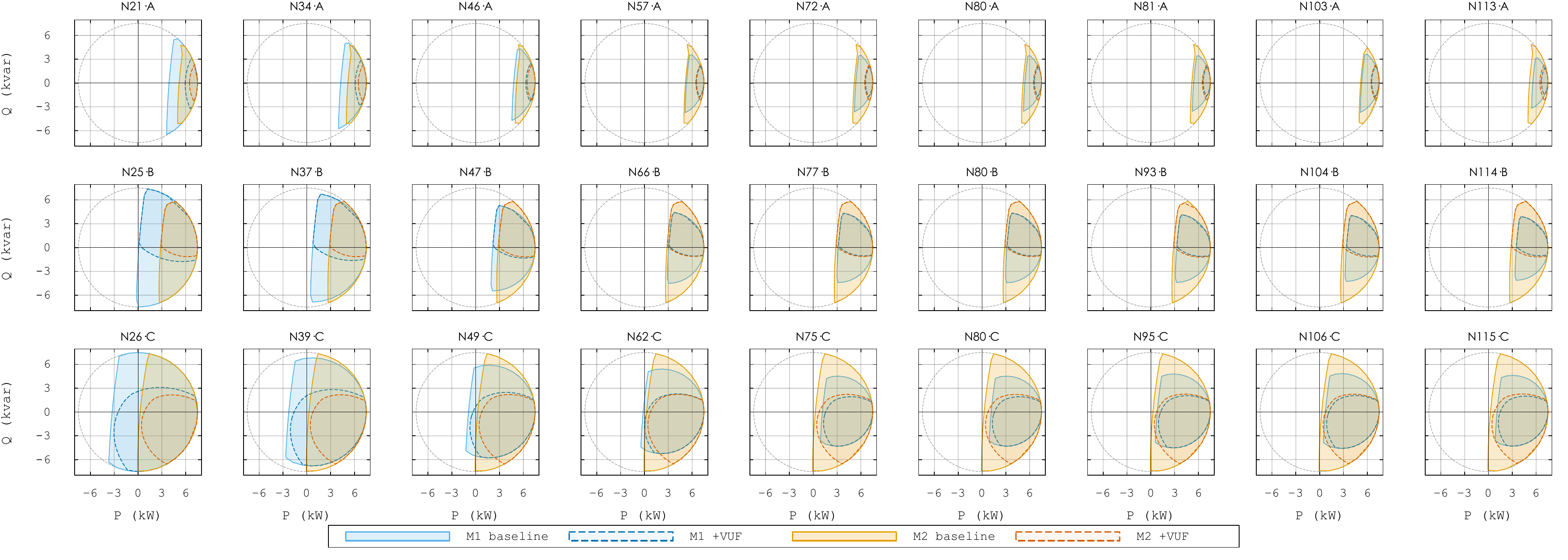}
    \caption{OE of Test Case 2 (Guide: M1:= Method~I, M2:= Method~II).}
    \label{fig:OE2}
\end{figure*}

The resulting OEs are shown in
Fig.~\ref{fig:OE1} for the 13-node feeder and in
Fig.~\ref{fig:OE2} for the larger LV feeder. In both cases, the
introduction of VUF limits reduces the feasible operating region,
although the magnitude and direction of this reduction vary across
phases and nodes. The figures also show that the differences
between Method~I (M1) and Method~II (M2) are system- and node-dependent: in some
subplots, the two methods lead to similar envelopes, while in
others they produce visibly different feasible regions. Rather
than interpreting these differences in detail here, the next
section examines their physical and operational meaning, with
particular focus on the role of voltage unbalance and on the
practical trade-off between the two fairness mechanisms.

\section{Discussion and Comparative Analysis}
\label{sec:discussion}

This section discusses the OEs in
Figs.~\ref{fig:OE1} and~\ref{fig:OE2}.
First, the effect of the VUF constraint is examined.
Then, M1 and M2 are compared, and the phase-dependent
structure of the envelopes is analyzed.
Finally, the main implications are summarized.

\subsection{Effect of the VUF Constraint}

Across both test cases, incorporating the VUF constraint
reduces the OEs of the DER units, although the degree of
reduction depends strongly on the phase to which each unit
is connected.
The reduction is not uniform in the P--Q plane either:
it is concentrated in the direction of maximum real power
export, where additional injection on an already unbalanced
phase most strongly amplifies the negative-sequence
voltage component.
The curtailment and reactive power directions are
comparatively less affected.

In Test Case~1, the effect is most pronounced on Phase~A
and Phase~B, which carry the heaviest and intermediate
loads, respectively.
At baseline, these phases already operate closer to
the VUF limit, so the constraint becomes active earlier
during the angular tracing and clips a larger portion of
the envelope boundary.
For Phase~C, the least loaded phase, the dashed and solid
boundaries are nearly coincident at all three nodes,
indicating that Phase~C injections have negligible influence
on voltage unbalance at the baseline operating point.

In Test Case~2, with more DER units and higher aggregate
penetration, the baseline operating point already exhibits
a higher unbalance level, and the VUF constraint
binds noticeably on all three phases.
The overall reduction in envelope area is larger
than in Test Case~1, and no phase is exempt from the effect.
This suggests that as DER penetration grows, the VUF
constraint transitions from a phase-selective restriction
to a network-wide limiting factor that must be accounted
for in OE computation regardless of the phase to which a
unit is connected.

\subsection{Comparison of M1 and M2 Across the Feeder}

The relative performance of the two methods depends on
where a unit is located in the network.
Rather than one method being uniformly better, the figures
reveal a consistent spatial gradient along the feeder.

Units located close to the substation tend to receive
larger envelopes under M1 than under M2.
These units have a low voltage impact on the network and
therefore receive high sensitivity weights under M1,
which translates directly into a large share of the
phase-wide headroom.
Under M2, the same units receive a more moderate
allocation, because the lexicographic procedure does not
use predetermined ratios but determines each unit's share
from the actual network feasibility at each stage.
In Test Case~1 this is most visible at Node~5 on Phase~A
and Node~6 on Phase~B.
In Test Case~2 the same pattern appears at nodes 21 and
34 on Phase~A, 25 and 37 on Phase~B, and 26 and 39 on Phase~C.

Toward the middle of the feeder, the two methods converge.
At these intermediate nodes, the weights assigned by M1
are a reasonable approximation of the allocation that M2
produces through its staged procedure, so the resulting
envelopes are nearly identical, as seen at Node~8 on
Phase~A and Node~9 on Phase~B in Test Case~1, and at
several mid-feeder nodes in Test Case~2.

At the far end of the feeder the situation reverses.
Deep-feeder units have a high voltage impact and
therefore receive small weights under M1, which caps
their headroom at a small fraction of the phase-wide
allocation.
Under M2, these units are not subject to a predetermined
cap; the staged procedure allocates them whatever headroom
the network can support, independently of their sensitivity.
As a result, M2 envelopes are larger than M1 at the far
end of the feeder, as seen at Nodes~11 and~12 in Test
Case~1 and at nodes 113, 114, and 115 in Test Case~2.

This spatial gradient reflects a fundamental characteristic
of proportional fairness in radial networks: M1 favors
near-slack units in proportion to baseline sensitivities,
whereas M2 derives the allocation from the network physics
at each stage, giving deep-feeder units more headroom and
near-slack units moderately less.
M2 is therefore the appropriate choice when equitable
treatment regardless of feeder position is the priority,
whereas M1 provides a practical and scalable alternative
when computation speed matters and a weight-based
allocation is acceptable.

\subsection{Phase-Dependent Envelope Structure}

The figures also show a consistent trend in how envelope
shapes change across phases.
Phase~A envelopes are anchored in the right half-plane,
with real power export as the dominant direction of
available headroom.
The curtailment direction is partially constrained by
the lower voltage bound on this most heavily loaded phase,
which is already near its limit at the baseline dispatch.
Phase~B envelopes shift downward, with more headroom
in the inductive (absorption) direction than in the
capacitive (injection) direction, reflecting the
intermediate voltage sensitivity of this phase.
Phase~C envelopes are by far the largest in both test
cases, approaching the full inverter capability circle
at several nodes, which confirms that network constraints
are nearly non-binding for the least loaded phase.

This phase-dependent structure is reproduced by both
M1 and M2, which confirms that it is a property of the
network's operating point rather than an artifact of
the fairness formulation.

\subsection{Summary of Implications}

The VUF constraint cannot be ignored in OE computation,
particularly as DER penetration increases.
Its effect is phase-selective in Test Case~1, but in Test
Case~2 it affects all phases and reduces envelope areas
to a degree that would be consequential in real operation.
Formulations that omit voltage unbalance therefore
systematically overestimate the admissible operating region.

An important and perhaps unexpected finding is that
the VUF constraint significantly narrows the gap between
M1 and M2.
Without VUF, the two methods produce noticeably different
envelopes at near-slack and deep-feeder nodes.
Once the VUF constraint is enforced, the larger envelope
of each unit is clipped more aggressively, since it extends
further into the regions where voltage unbalance becomes
problematic.
As a result, the VUF-constrained envelopes of M1 and M2
are considerably closer to each other than their
unconstrained counterparts, as visible from the dashed
boundaries in both figures.

The price of fairness in Table~\ref{tab:pof} shows that
VUF constraints also make fairness more costly.
Without VUF, the binding constraints are mainly local
voltage limits, and M2 allocates the available headroom
with a smaller loss of envelope area than M1.
Once VUF constraints are enforced, the utilitarian
benchmark can still concentrate headroom on the units
whose injections least increase the negative-sequence
voltage, whereas both fair methods move all units on a
phase together and are therefore limited by the most
unbalance-sensitive unit.
Consequently, the PoF of both methods increases and the
difference between them largely vanishes in both networks,
confirming in aggregate terms the convergence observed in
the envelope shapes.
The PoF is also markedly higher in NW2 than in NW1,
consistent with the larger number of units sharing each
phase, which gives the benchmark more heterogeneity to
exploit.

This has a direct implication for the choice of method.
M2 provides the strongest theoretical fairness guarantee:
no unit's headroom can be improved without reducing that
of a less well-served unit.
However, when VUF constraints are active, M1 produces
envelopes that are nearly as fair as M2 in practice, at a
comparable price of fairness, while requiring a single
optimization solve per angle and approximately one-sixth
of the computation time of M2 in Test Case~2.
M1 is therefore attractive for operational settings in
which computational speed is a priority, whereas M2 remains
preferable for offline planning and regulatory assessments
where the strongest fairness guarantee is required.

\subsection{Special Considerations}
\label{sec:ES-Cons}

\subsubsection{Phase Interdependence and Baseline Operating Point}
Although the {OE} computation is performed independently for
each phase, the phases remain electrically coupled through
neutral voltage shifts and mutual impedances.
Consequently, the units connected to the phases not under
analysis must be assigned fixed operating points. Three
assumptions of decreasing physical fidelity are possible:
\begin{itemize}
    \item \textbf{Baseline {OPF} (adopted in this work):}
    units on non-analyzed phases are fixed at the dispatch
    obtained from a baseline {OPF}, which provides the most
    accurate representation of inter-phase interactions
    under normal operating conditions.
    \item \textbf{Nominal values:} units on non-analyzed phases
    are set to their rated power. This option suits cases in
    which the {DSO} has incomplete network information, as it
    partially compensates for data gaps while reducing the
    computational burden.
    \item \textbf{Full curtailment ($P = Q = 0$):} units
    on non-analyzed phases are assumed to be de-energized.
    This permits computation under extreme data scarcity but
    effectively decouples the phases, approaching a
    single-phase {OPF} and considerably overestimating the
    available network headroom.
\end{itemize}
The mathematical formulation is independent of the baseline
assumption and yields valid results under all three options.
The baseline {OPF} assumption is adopted throughout this work
because it provides the tightest and most operationally
representative envelopes. A systematic quantification of the
sensitivity of {OEs} to this choice, particularly under
partial or unavailable network data, is left for future work.

\subsubsection{Operational Validity of Computed {OEs}}
When a unit produces zero output, for example a {PV} unit
during periods of low or zero irradiance, the operating point
($P = Q = 0$) lies within the computed {OE} boundary and
therefore satisfies the envelope constraints. The {OE} represents a boundary of permissible operation rather
than a prescribed setpoint. Note, however, that the envelopes are
traced along coordinated directions from the baseline, with units
on the non-analyzed phases fixed at their baseline dispatch.
Network feasibility of arbitrary, independently chosen operating
points within each envelope is therefore not formally guaranteed,
and real-time deviations may still give rise to phase imbalance.
This limitation can be addressed by complementing the OEs with
real-time controllers, such as online feedback
optimization~\cite{Hauswirth2024}, which steer the realized
operating point toward compliance with voltage and VUF limits.

\section{Conclusion}
\label{sec:conclusion}

This paper developed a unified framework for computing voltage unbalance-aware P--Q operating envelopes for PV inverters in unbalanced three-phase distribution networks. The framework embeds nodal VUF constraints, together with voltage magnitude limits and inverter capability constraints, in a three-phase AC-OPF formulation, and traces the boundary of the feasible P--Q region through a directional OPF sweep. Two flexibility allocation methods are incorporated: M1, a network-aware proportional fairness mechanism based on voltage sensitivities to power injections, and M2, lexicographic max--min fairness.

Case studies on two unbalanced test feeders showed that VUF constraints shrink and reshape the feasible P--Q regions. These effects differ across phases and nodes and are most pronounced where the unconstrained regions are broad. Voltage unbalance can therefore be a binding constraint on DER flexibility and should be explicitly considered when computing operating envelopes.

Enforcing VUF constraints also brings the regions obtained under M1 and M2 closer together, although notable differences persist at some nodes. Power quality requirements thus affect not only the available flexibility but also how much the choice of fairness method matters. Relative to a utilitarian benchmark, enforcing VUF constraints increases the price of fairness, measured as the loss of envelope area, for both methods and reduces the difference between them. M1 required roughly one-sixth of the computation time of M2, which makes it attractive for scalable flexibility allocation. M2 remains valuable when the least-favored DER must be prioritized or when a more detailed fairness assessment is needed.

Since OEs bound permissible operation but do not determine the
realized injections, future work will investigate real-time
controllers, such as online feedback optimization, that keep the
realized operating point within voltage and VUF limits.

\balance
\section*{Acknowledgment}

The work of Alireza Zabihi was supported by the 2023 FPI-UPM call for Predoctoral Contracts within the framework of the 2021-2023 State Plan for Scientific, Technical, and Innovative Research.

\ifCLASSOPTIONcaptionsoff
  \newpage
\fi

\bibliographystyle{IEEEtran}
\balance
\bibliography{Bibliography}
\end{document}